\documentclass[journal=jacsat,manuscript = communication]{achemso}

\usepackage{graphicx}
\usepackage{amssymb}
\usepackage{multirow}
\usepackage{booktabs}
\usepackage{bm}
\setkeys{acs}{articletitle = true}

\usepackage[normalem]{ulem}
\usepackage[usenames,dvipsnames]{xcolor} 

\newif\ifASHighlitedChanges
\def\ifASHighlitedChanges{\iftrue}
\ifASHighlitedChanges
  
  \def\STRIKEAS#1{{\color{Orange}\sout{#1}}}
\else
  
  \def\STRIKEAS#1{\relax}
\fi

\newif\ifRPSHighlitedChanges
\def\ifRPSHighlitedChanges{\iffalse}
\ifRPSHighlitedChanges
  \def\EDITSRPS#1{{\color{DarkOrchid}#1}}
  \def\STRIKERPS#1{{\color{DarkOrchid}\sout{#1}}}
\else
  \def\EDITSRPS#1{#1}
  \def\STRIKERPS#1{\relax}
\fi

\newif\ifJCCHighlitedChanges
\def\ifJCCHighlitedChanges{\iftrue}
\ifJCCHighlitedChanges
  
  \def\STRIKEJCC#1{{\color{blue}\sout{#1}}}
\else
  
  \def\STRIKEJCC#1{\relax}
\fi

\newif\ifRKHighlitedChanges
\def\ifRKHighlitedChanges{\iftrue}
\ifRKHighlitedChanges
  
  \def\STRIKERK#1{{\color{forestgreen}\sout{#1}}}
\else
  
  \def\STRIKERK#1{\relax}
\fi

\title{Soft interactions modify the diffusive dynamics of polymer-grafted nanoparticles in solutions of free polymer}

\author{Ryan Poling-Skutvik}
\affiliation{Department of Chemical and Biomolecular Engineering, University of Houston, Houston, TX 77204-4004}

\author{Ali H. Slim}
\affiliation{Department of Chemical and Biomolecular Engineering, University of Houston, Houston, TX 77204-4004}

\author{Suresh Narayanan}
\affiliation{Advanced Photon Source, Argonne National Laboratory, Argonne, IL 60439}

\author{Jacinta C. Conrad}
\email{jcconrad@uh.edu}
\affiliation{Department of Chemical and Biomolecular Engineering, University of Houston, Houston, TX 77204-4004}

\author{Ramanan Krishnamoorti}
\email{ramanan@uh.edu}
\affiliation{Department of Chemical and Biomolecular Engineering, University of Houston, Houston, TX 77204-4004}

\date{\today}

\begin{document}

\begin{abstract}

We examine the dynamics of silica particles grafted with high molecular weight polystyrene suspended in semidilute solutions of chemically similar linear polymer using x-ray photon correlation spectroscopy. The particle dynamics decouple from the bulk viscosity despite their large hydrodynamic size and instead experience an effective viscosity that depends on the molecular weight of the free polymer chains. Unlike for hard sphere nanoparticles in semidilute polymer solutions, the diffusivities of the polymer-grafted nanoparticles do not collapse onto a master curve as a function of normalized length scales. These results suggest that the soft interaction potential between polymer-grafted nanoparticles and free polymer allows polymer-grafted nanoparticles to diffuse faster than predicted based on bulk rheology and modifies the coupling between grafted particle dynamics and the relaxations of the surrounding free polymer.

\end{abstract}

\section{Introduction}

Attaching polymers to surfaces modifies the interactions between nanoparticles and surrounding environments. Such fine-tuning of nanoparticle interactions is important to improve the biocompatibility of targeted drug delivery vectors,\cite{Gref1994,Owens2006,Tong2012,Kreyling2015} control self-assembled structures in nanocomposites,\cite{Akcora2009,Chevigny2011,Kumar2013,Martin2015} or stabilize emulsions.\cite{Alvarez2012,Foster2014,Kim2015} For these applications, the efficacy of polymer-grafted nanoparticles (PGNPs) requires that the particles remain stable and transport effectively when dispersed into complex fluids. Whereas the long-time dynamics of large hard sphere colloids through complex fluids are well understood, multiple factors complicate predictions of the motion of PGNPs. First, PGNPs are often comparably sized to heterogeneities in the surrounding complex fluid, violating an assumption underlying microrheology theory.\cite{Mason1995,Mason2000,Squires2010} Second, PGNPs are soft particles whose `softness' can be characterized by their elastic deformability\cite{Philippe2018} or through the steepness and range of their repulsive interactions.\cite{Koumakis2012,Zhou2014,Schneider2017} The combination of soft interactions between grafted polymers and hard interactions of the nanoparticle cores leads to elastic moduli and yield stresses for PGNP suspensions lower than those of hard sphere colloids and higher than those of ``ultra-soft'' star-like polymers or micelles.\cite{Erwin2010,Koumakis2012,Gupta2015} Finally, the tethering of polymer to the particle surface significantly changes the grafted polymer relaxations\cite{Yakubov2004,Frielinghaus2010,Agarwal2012,Poling-Skutvik2017} and may therefore affect the transport of PGNPs.

Here, we investigate the dynamics of silica nanoparticles grafted with high molecular weight polystyrene so that the grafted polymer and silica core are comparably sized. The PGNPs are dispersed into solutions of free polystyrene and the dynamics of the PGNP center-of-mass are probed using x-ray photon correlation spectroscopy (XPCS). The PGNP diffusivity systematically depends on the free polymer molecular weight, diffusing faster in solutions with the same bulk viscosity but higher molecular weight. Although similar dependences have been observed for hard sphere nanoparticles, the PGNPs are much larger than the length scale at which hard spheres begin to decouple from bulk viscosity and the PGNP diffusivity does not collapse according to relative size as it does for hard spheres. We propose that these unique transport properties of PGNPs arise from the soft interaction between the PGNP corona and the free chains in solution. 

\section{Materials and Methods}

We graft\cite{Poling-Skutvik2017} high molecular weight M$_\mathrm{W}$ = 355 kDa polystyrene onto silica nanoparticles with a radius $R = 24$ nm \EDITSRPS{ to form a PGNP with a morphology intermediate between a pseudo-hard sphere (low grafted $M_\mathrm{W}$) and a star polymer (high grafted $M_\mathrm{W}$).} Due to sample preparation constraints, two separate batches of grafted particles are synthesized -- one for X-ray and neutron scattering experiments with grafting density $\sigma = 0.054 \pm 0.015$ \EDITSRPS{chains nm$^{-2}$} and one for rheology experiments with $\sigma = 0.069 \pm 0.015$ \EDITSRPS{chains nm$^{-2}$}. Particles from both batches have similar hydrodynamic radius $R_\mathrm{H} = 110$ nm determined from dynamic light scattering and overlap concentration $c^*_\mathrm{PGNP} = 0.023$ g mL$^{-1}$ determined from intrinsic viscosity. Small-angle neutron scattering (SANS) experiments are conducted on the NG7 30m beamline\cite{Glinka1998,Kline2006} at the NIST Center for Neutron Research on PGNPs disperse in partially deuterated d$_5$-2-butanone and partially deuterated d$_3$-polystyrene with $M_\mathrm{W} = 140$, 640, and 1100 kDa to contrast-match the silica core. XPCS measurements were conducted on the 8-ID-I beamline at Argonne National Lab on PGNPs dispersed in \emph{protonated} solvent and polystrene with $M_\mathrm{W} = 150$, 590, \EDITSRPS{ 1100, and 15000} kDa, which span the molecular weight of the grafted chain over a range that controls PGNP dispersion in polymer melts.\cite{Kumar2013} \EDITSRPS{PGNPs remained well dispersed in all solutions (Supporting Information).} Steady-shear rheology experiments on protonated solutions were conducted on a Discovery Hybrid Rheometer (TA Instruments, HR-2) using a Couette geometry.

\section{Results and Discussion}
\begin{figure}[htb!]
\includegraphics[width = 3.25 in]{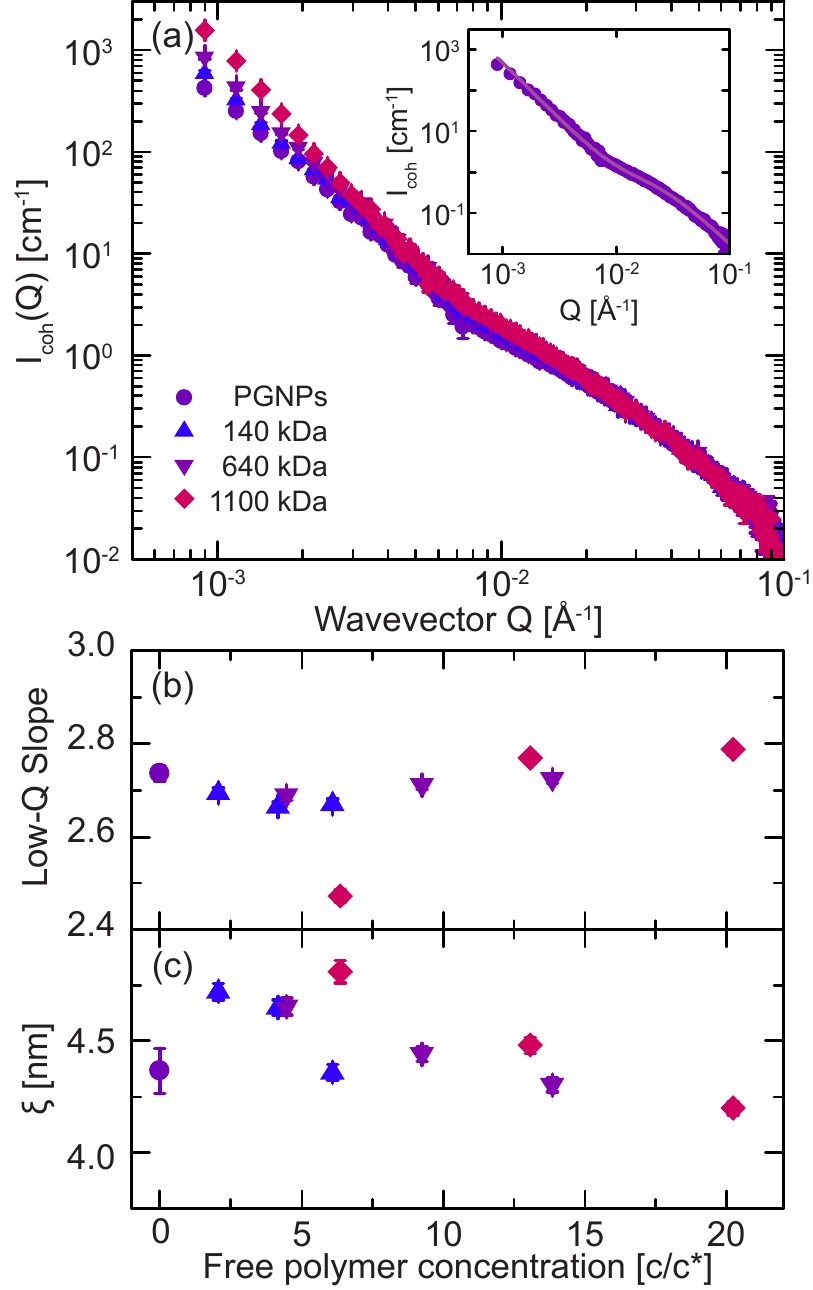}
\caption{\label{fig:SANS}(a) Coherent SANS intensity $I_\mathrm{coh}$ as a function of wavevector $Q$ for PGNPs dispersed at $\approx 1c^*_\mathrm{PGNP}$ in solutions of d$_5$-2-butanone and free d$_3$-polystyrene of various $M_\mathrm{W}$ at $c = 0.15$ g mL$^{-1}$. \textit{Inset:} $I_\mathrm{coh}(Q)$ for PGNPs with no free polymer. Solid curve is best fit to Eq.\ \ref{eq:SANS}. (b) Low-$Q$ slope $m$ and (c) correlation length $\xi$ as a function of free polymer concentration.}
\end{figure}

Previous experiments\cite{Stiakakis2002,Stiakakis2005} and simulations\cite{Camargo2010,Camargo2012} have identified rich structural changes for PGNPs dispersed in solutions of free chains. We first investigate using SANS the structure of PGNPs dispersed in partially deuterated solutions to isolate scattering from the grafted polymer (Fig.\ \ref{fig:SANS}). At large $Q$, the intensity derives from intra- and interchain correlations whereas at low $Q$, the scaling of the intensity corresponds to the sharpness of the interface between the PGNP corona and surrounding solvent. Because these PGNPs are large, we do not observe a low-$Q$ plateau and thus cannot model the scattering intensity using previously described models derived for grafted morphologies.\cite{Pedersen1996,Pedersen2003,Foster2011,Hore2013} Instead, we model the full scattering intensity as the sum of a Lorentzian to capture the polymer conformation inside the corona and a power law to model the sharpness of the corona-solvent interface according to
\begin{equation}
\label{eq:SANS}
I_\mathrm{coh}(Q) = \frac{I_\mathrm{poly}}{1+(Q\xi)^{1/\nu}} + AQ^{-m},
\end{equation}
where $I_\mathrm{poly}$ is corona polymer intensity, $\xi$ is the correlation length between grafted chains, $\nu$ is the excluded volume parameter, and $m$ is the low-$Q$ slope. Because the amount of grafted polymer does not change in these solutions, we hold $I_\mathrm{poly}$ constant and let the other parameters float.

Physically, grafted brushes are expected to compress in solutions of free chains.\cite{Stiakakis2002, Stiakakis2005, Wilk2010, Poling-Skutvik2017} Changes in the scattering pattern as the concentration and molecular weight of free polymer varies are consistent with this physical picture. When polymer is first added to the system, $m$ decreases and $\xi$ increases (Fig.\ \ref{fig:SANS}(b,c)). These changes could be caused by a small expansion of the grafted corona. Upon increasing free polymer concentration, $m$ decreases in solutions with low concentrations of low $M_\mathrm{W}$ free polymer but increases in more concentrated solutions of higher $M_\mathrm{W}$. This dependence suggests that the boundary between the grafted corona and surrounding solution becomes slightly more diffuse at low concentrations and sharper at higher concentrations. Inside the grafted corona, $\xi$ decreases with increasing free polymer concentration as the corona compresses and decreases more strongly in solutions with lower molecular weight. This compression is caused by an increase in the solution osmotic pressure by the free chains and a logarithmic interaction potential between free chains and PGNPs\cite{Mayer2007,Camargo2010,Camargo2012} that is softer than the power-law interactions between hard-sphere colloids and polymers in solution.\cite{Asakura1958,Bolhuis2003,Kleshchanok2008}

\begin{figure}[htb!]
\includegraphics[width = 3.25 in]{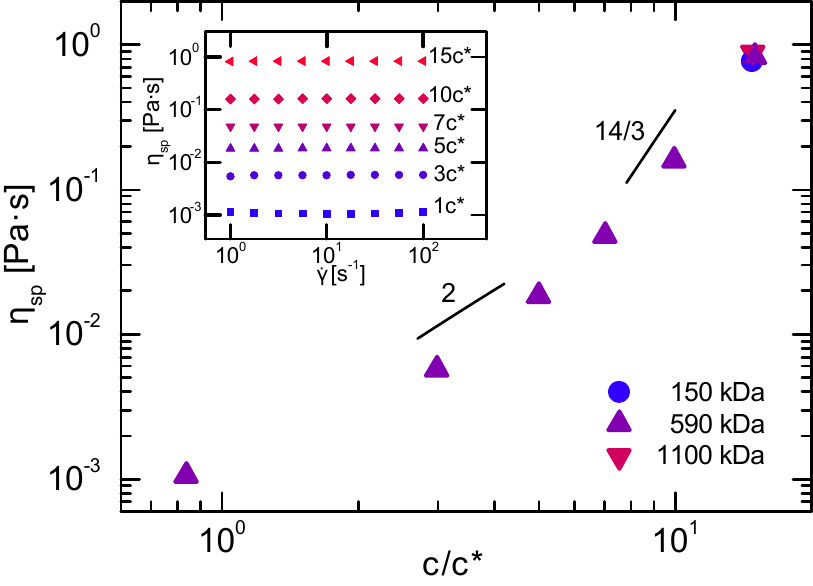}
\caption{\label{fig:Viscosity} Specific viscosity $\eta_\mathrm{SP} = \eta - \eta_0$ as a function of free polystyrene concentration $c/c^*$ for solutions of varying molecular weight. \textit{Inset:} Specific viscosity $\eta_\mathrm{SP}$ for 590 kDa solutions as a function of shear rate $\dot{\gamma}$. All samples have a grafted polystyrene concentration of 0.5$c_\mathrm{PGNP}^*$.}
\end{figure}

After determining that the structure of the PGNPs agrees with previous studies, we investigate whether the low PGNP concentration modifies bulk solution properties. For polymer solutions, bulk viscosity $\eta$ scales with the relative concentration $c/c^*$, independent of $M_\mathrm{W}$. To verify that this scaling collapse holds for solutions containing free polymer as well as PGNPs, we measure the steady shear viscosity as a function of shear rate $\dot{\gamma}$, free polymer concentration, and free polymer molecular weight (Fig.\ \ref{fig:Viscosity}). The viscosity is Newtonian with relaxation times $\lesssim 1$ ms (inset to Fig.\ \ref{fig:Viscosity}) and scales as predicted with $\eta \sim \left(c/c^*\right)^2$ when unentangled and as $\eta \sim \left(c/c^*\right)^{14/3}$ when entanglements dominate.\cite{Rubinstein2003}. Moreover, the measured viscosities of 15$c^*$ solutions exhibit no significant changes as a function of $M_\mathrm{W}$. These rheological characteristics are in excellent agreement with theory for solutions of free polymers, indicating that PGNPs do not perturb the bulk solution rheology.

\begin{figure}[htb!]
\includegraphics[width = 3.25 in]{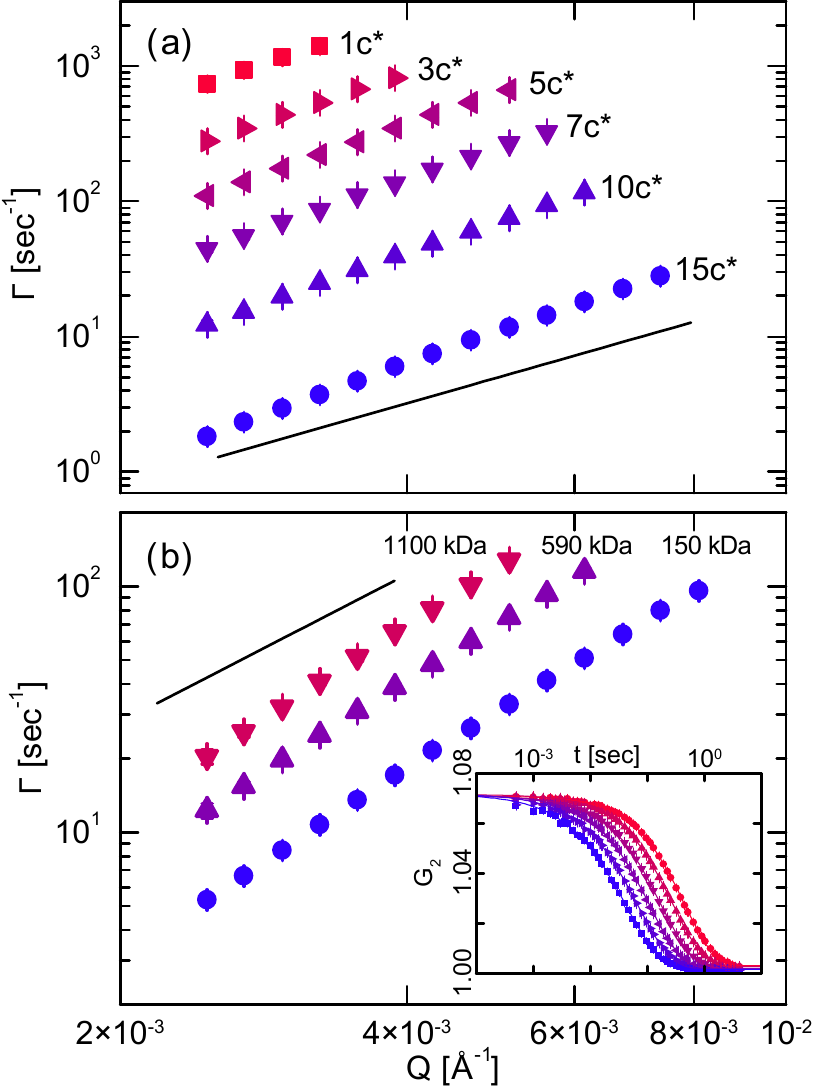}
\caption{\label{fig:Gamma} Relaxation rate $\Gamma$ from XPCS as a function of wavevector $Q$ at (a) various concentrations of 590 kDa free polystyrene and at (b) $c/c^* = 10$ for $M_\mathrm{W} = 150$, 590, and 1100 kDa. All samples have a PGNP concentration of 0.5$c_\mathrm{PGNP}^*$. Solid lines indicate $Q^{2}$ scaling. \textit{Inset}: Representative intensity autocorrelation function G$_2$ as a function of lag time for \mbox{0.003 \AA$^{-1}$} (red) $\leq Q \leq$ \mbox{0.0074 \AA$^{-1}$} (blue) for a solution of $c/c^* = 23$ and $M_\mathrm{W}$ = 1100 kDa. Solid curves are stretched exponential fits.}
\end{figure}

Having confirmed that the structure and solution rheology follow the expected behavior, we investigate the dynamics of the PGNPs in the polymer solutions using XPCS. Because the x-ray scattering is dominated by the silica core, the measured dynamics represent only those of the center-of-mass of the PGNP and do not directly reflect the relaxations of the grafted brushes. The intensity autocorrelation curves $G_2$ decay faster with time at smaller wavevectors $Q$ and are well fit by $G_2(Q,\Delta t) = 1 + BG_1(Q,\Delta t)^2 + \varepsilon$ where $B$ is the Siegert factor that depends on experimental geometry, $G_1(Q,\Delta t) = \exp\left[ - \left( \Gamma \Delta t \right)^\beta \right]$ is the field correlation function, $\beta \approx 0.85$ is a stretching exponent that arises from the polydisperse size of the PGNPs, and $\varepsilon$ captures any residual noise (inset to Fig.\ \ref{fig:Gamma}). \EDITSRPS{The relaxation rate follows $\Gamma = DQ^2$, where $D$ is the PGNP diffusivity} (Fig.\ \ref{fig:Gamma}). Because the free polymer increases the bulk solution viscosity, the particle dynamics slow with increasing $c/c^*$. Surprisingly, there is an additional dependence on the free polymer $M_\mathrm{W}$: PGNPs diffuse more slowly in solutions of low $M_\mathrm{W}$ than in solutions of high $M_\mathrm{W}$ at the same $c/c^*$ (Fig.\ \ref{fig:Viscosity}). This $M_\mathrm{W}$-dependence indicates that the nanoscale dynamics are fundamentally different from those on the macroscale.

\begin{figure}[htb!]
\includegraphics[width = 3.25 in]{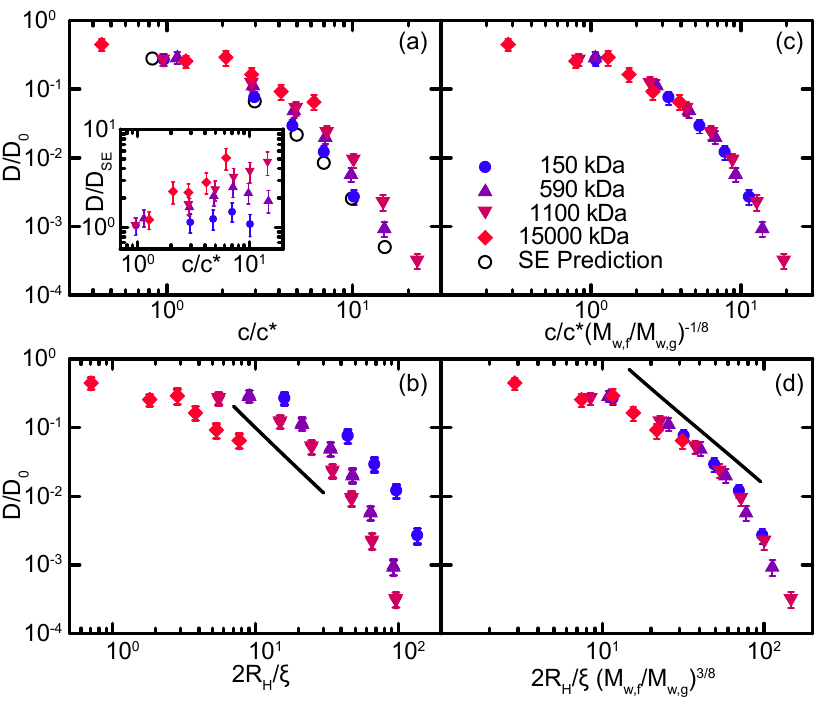}
\caption{\label{fig:Diffusivities} Normalized PGNP diffusivity $D/D_0$ as a function of (a) free polymer concentration $c/c^*$, (b) the ratio of PGNP size and correlation length $R_\mathrm{H}/\xi$, and M$_\mathrm{W}$-modified (c) concentration and (d) length scale ratio. \textit{Inset:} Diffusivity normalized to SE predictions $D/D_\mathrm{SE}$ as a function of $c/c^*$. Open symbols in (a) represent the expected diffusivity based on bulk rheology (Fig.\ \ref{fig:Viscosity}). Solid lines represents predicted scaling of -2 for hard spheres.\cite{Cai2011} All samples have a PGNP concentration of 0.5$c_\mathrm{PGNP}^*$.}
\end{figure}

To quantify this difference between the nanoscale dynamics of PGNPs and microrheological predictions based on bulk viscosity, we analyze how the PGNP diffusivity changes with free polymer concentration and molecular weight (Fig.\ \ref{fig:Diffusivities}(a)). Specifically, we compare the measured PGNP diffusivity to predictions based on bulk viscosity $\eta$ using the Stokes-Einstein (SE) expression $D_\mathrm{SE} = k_\mathrm{B}T/6\pi\eta R_\mathrm{H}$. For \EDITSRPS{$c/c^* \approx 1$, $D/D_\mathrm{SE} \approx 1$} for all free polymer molecular weights. As \EDITSRPS{$c/c^*$} increases, \EDITSRPS{$D/D_\mathrm{SE} \approx 1$} in the 150 kDa solutions \EDITSRPS{but increases with increasing $c/c^*$ and $M_\mathrm{W}$} (inset to Fig.\ \ref{fig:Diffusivities}(a)). \EDITSRPS{Although the PGNPs compress in solutions, this compression cannot explain the observed discrepancy in $D/D_\mathrm{SE}$. According to SANS (Fig.\ \ref{fig:SANS}) and theoretical predictions\cite{Truzzolillo2011} (Supporting Information), the PGNPs in low $M_\mathrm{W}$ solutions shrink more and should therefore diffuse faster. The observed behavior is exactly opposite: PGNPs diffuse \emph{faster} at high $M_\mathrm{W}$. Instead, the discrepancy in $D/D_\mathrm{SE}$ must originate from interactions between PGNPs and the surrounding polymer solution on the nano- or microscale.}

The dynamics of large colloidal particles through complex fluids are excellently described in the microrheological framework by the SE or generalized SE expressions.\cite{Mason1995,Mason2000,Squires2010} Nanoparticles, however, often diffuse faster than expected \EDITSRPS{because they} are comparably sized to \EDITSRPS{the polymer coils in solution}.\cite{Cai2011} For these PGNPs, however, the hydrodynamic radius $R_\mathrm{H} \approx 110$ nm is much larger than the \EDITSRPS{polymer} radius of gyration $R_\mathrm{g} = 13$, 27, and 38 nm for $M_\mathrm{W} = 150$, 590, and 1100 kDa, respectively.\cite{Wagner1985} Based on the relative size of the PGNPs it is surprising to observe the $M_\mathrm{W}$-dependence shown in Fig.\ \ref{fig:Diffusivities}(a). A recent \EDITSRPS{coupling} theory attempts to explain the enhanced diffusion of hard sphere nanoparticles in semidilute polymer solutions \EDITSRPS{through a length scale ratio via $D/D_0 \sim (R_\mathrm{H}/\xi)^{-2}$, where $D_0$ is the nanoparticle diffusivity in solvent}.\cite{Cai2011} This theory excellently collapses diffusivities measured experimentally\cite{Poling-Skutvik2015} and in simulations\cite{Chen2018} for hard-sphere nanoparticles \EDITSRPS{but cannot collapse the PGNP dynamics} (Fig.\ \ref{fig:Diffusivities}(b)). The significant difference between the dynamics of PGNPs in solutions with varying $M_\mathrm{W}$ \EDITSRPS{indicates that PGNPs experience local heterogeneities in semidilute polymer solutions differently than hard-sphere particles do.} 

\EDITSRPS{The ratio between free and grafted polymer $M_\mathrm{W}$ has been shown to control the morphology of PGNPs\cite{Akcora2009,Kumar2013} and star polymers\cite{Stiakakis2005} in polymer melts and solutions. To determine if this ratio also affects dynamics,} we plot the PGNP diffusivity as a function of $c/c^*$ and $R_\mathrm{H}/\xi$ modified by the ratio of free to grafted polymer molecular weights $M_\mathrm{W,f}/M_\mathrm{W,g}$ (Fig.\ \ref{fig:Diffusivities}(c,d)). \EDITSRPS{Incorporating the $M_\mathrm{W}$ ratio cleanly collapses the PGNP diffusivities across \emph{two orders of magnitude} in both $c/c^*$ and free polymer $M_\mathrm{W}$.} In these collapses, the molecular weight scaling exponents are empirically determined to minimize spread in the data. The inverse scaling of $(M_\mathrm{W,f}/M_\mathrm{W,g})^{-1/8}$ required to collapse diffusivity as a function of $c/c^*$ indicates that the PGNPs experience a lower effective viscosity in solutions of higher $M_\mathrm{W,f}$. By contrast, the collapse according to $R_\mathrm{H}/\xi$ requires a direct scaling with $(M_\mathrm{W,f}/M_\mathrm{W,g})^{3/8}$ to offset the overcorrection from the hard-sphere scaling. \EDITSRPS{Although the numerical exponents may vary with grafting density, core size, or grafted $M_\mathrm{W}$, the collapse indicates that the $M_\mathrm{W}$ ratio controls the nanoscale interactions that affect PGNP dynamics in polymer solutions}.

\EDITSRPS{Beyond the importance of the $M_\mathrm{W}$-ratio, the PGNP diffusivity also has a different functional dependence on $R_\mathrm{H}/\xi$ than hard spheres. Over a limited range ($20 \lesssim 2R_\mathrm{H}/\xi \lesssim 50$), PGNP diffusivity decreases similarly to theoretical predictions for hard spheres. At higher concentrations, however, PGNP diffusivity decreases more sharply than a power-law with a slope of -2, which suggests that the dynamics of PGNPs are slowed by entanglements between free polymers at high $c/c^*$ where bulk viscosity increases rapidly (Fig.\ \ref{fig:Viscosity}). Such entanglement-controlled dynamics are not observed for hard sphere nanoparticles until the particle radius exceeds the tube diameter in the entangled solution.\cite{Cai2011,Guo2012} Although entanglements between star polymer arms and free chains can significantly alter the diffusion of the star core, predictions based on this theory\cite{McLeish2002} cannot explain the $M_\mathrm{W}$-dependence observed here (Supporting Information). Thus, the diffusive dynamics of PGNPs through polymer solutions cannot be described through theories derived purely for hard sphere nanoparticles or star polymers.}

\EDITSRPS{To guide future investigations, we consider a variety of physical phenomena that may control how the dynamics of PGNPs differ from those hard spheres or star polymers. First, compression of PGNPs in solution cannot fully explain the observed dynamic phenomenon and may not follow star polymer predictions due to the finite size of the core. Complementary characterizations of PGNP dynamics and structure on long length scales are essential. Second, interactions between free chains and the grafted chains can affect the grafted polymer dynamics\cite{Poling-Skutvik2017} or the hydrodynamic drag on the PGNP surface\cite{Begam2015,Chandran2016}. Third, the deformability and dynamic relaxations of the grafted polymer may modify how PGNPs couple to segmental relaxations of the free polymer mesh and thus affect transport properties.\cite{Cai2011,Cai2015} Although assessing the magnitude of these different phenomena is beyond the scope of this letter, we have identified that soft interactions introduced by the grafted polymer prevents PGNPs from being treated simply as hard spheres or star polymers.
}

\section{Conclusion}
We investigated the dynamics of nanoparticles grafted with long polymer chains dispersed in solutions of free polymer. Although the PGNPs are much larger than the free polymer chains, their dynamics decouple from bulk viscosity in solutions of high molecular weight free polymer, diffusing up to five times faster than expected. Additionally, the PGNP dynamics depend on free polymer molecular weight and do not collapse according to predictions for hard spheres. This lack of collapse suggests that the PGNPs experience local heterogeneities and couple to relaxations in the surrounding fluid differently than hard spheres. We posit that these differences between PGNPs and hard spheres arises due to the soft interaction potential between the grafted corona and the free polymer. Many parameters -- including core size, grafted polymer molecular weight, and grafting density -- control the physical structure, organization,\cite{Akcora2009} and interaction potential of PGNPs. Understanding how these parameters modify transport properties is essential to controlling the efficacy of PGNPs dispersed in complex fluids.

\begin{acknowledgement}
We thank Megan Robertson for access to the rheometer. This research is supported by grants from the Welch Foundation (E-1869) and NSF (CBET-1704968). This research used resources of the Advanced Photon Source, a user facility operated for the DOE Office of Science by Argonne National Laboratory under Contract No. DE-AC02-06CH11357. We acknowledge the support of the National Institute of Standards and Technology, U.S. Department of Commerce, in providing the neutron research facilities used in this work.
\end{acknowledgement}

\textbf{Supporting information}
Supporting information contains SAXS curves, calculations of PGNP compression, and estimates of arm retraction dynamics.


\bibliography{biblio}

\end{document}


\subsection{Predicted compression of PGNPs in solutions of linear chains}
\begin{figure}[htb!]
\includegraphics[width = 2.5 in]{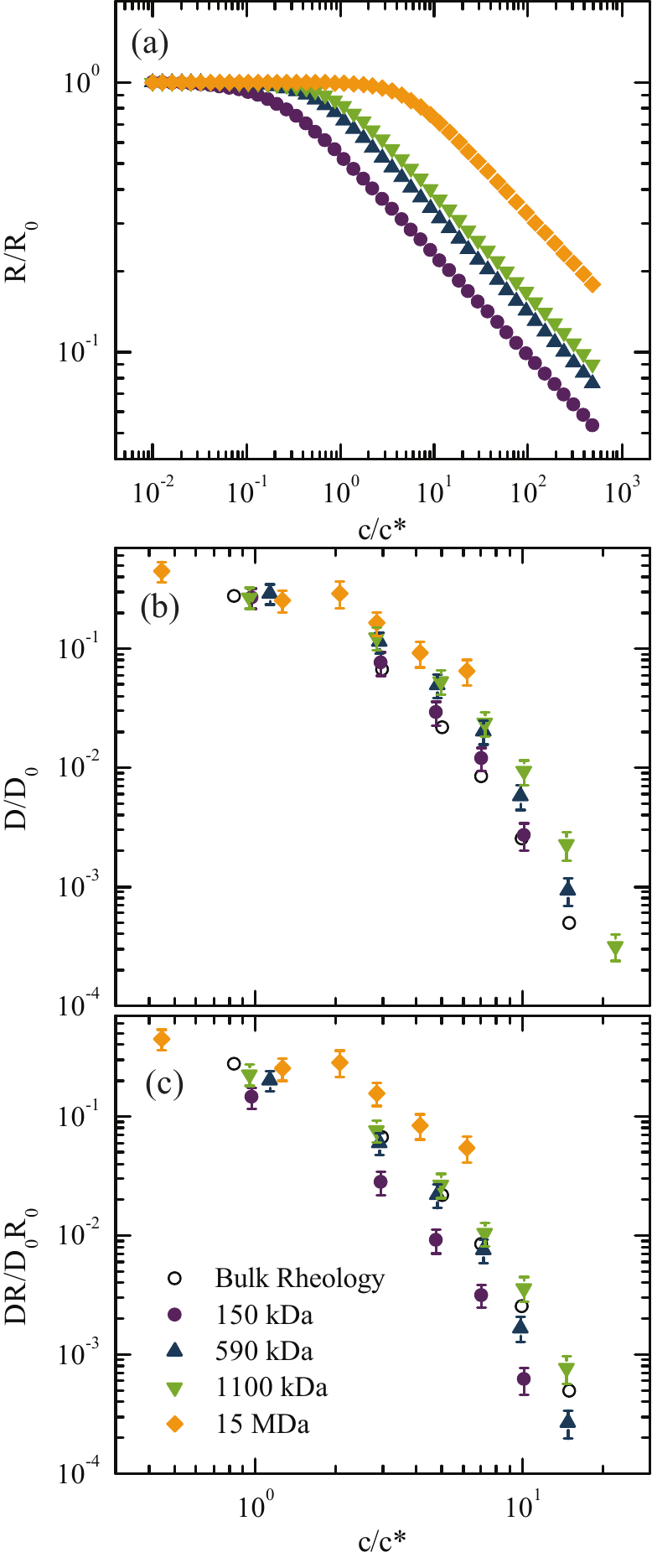}
\caption{\label{fig:Compression} (a) Predicted change in the size of PGNPs dispersed in solutions of linear polymer chains with varying $M_\mathrm{w}$ as a function of normalized concentration $c/c^*$ based on eqn.\ \ref{eqn:compression}. PGNP diffusivity normalized by (b) the hydrodynamic radius $R_\mathrm{H}$ in pure solvent and by (c) the predicted compressed size shown in (a). }
\end{figure}

The compression of star polymers in solutions of linear polymers was derived by balancing the elasticity of the star chains with the osmotic pressure of the solution, according to
\begin{equation}
\label{eqn:compression}
\frac{3faR_\mathrm{s}}{N_\mathrm{s}} + \frac{4\pi b^3 \Phi_\mathrm{L}[1+P(\Phi_\mathrm{L}]}{\delta^3 N_\mathrm{s}^{3\nu}f^{3/5}} - \frac{3\nu N_\mathrm{s}^2 f^2 a^3}{2R_\mathrm{s}^4} = 0,
\end{equation}
where 
\begin{equation}
P(x) = \frac{1}{2} x \exp\left\{ \frac{1}{4} \left[ \frac{1}{x}+\left(1-\frac{1}{x} \right) \ln (x+1) \right]\right\},
\end{equation}
\begin{equation}
x(\Phi_\mathrm{L}) = \frac{1}{2} \Phi_\mathrm{L} \pi^2 \left[ 1 + \frac{1}{4}\left( \ln 2 + \frac{1}{2} \right) \right],
\end{equation}
$\Phi_\mathrm{L} = c/c^*$ is the volume fraction of linear chains, $f$ is the functionality of the stars, $a$ is the monomer size, $N_\mathrm{s}$ is the degree of polymerization of the star polymer arms, $b = 1.3$ represents the penetrability of the star, $\delta = R_\mathrm{L}/R_\mathrm{s}$ is the size ratio of linear to star polymer, and $\nu$ is the Flory exponent.\cite{Truzzolillo2011} Because PGNPs geometrically resemble star polymers, we use eqn. \ref{eqn:compression} to estimate the compression of PGNPs dispersed into solutions of linear chains with varying molecular weights, as shown in Fig.\ \ref{fig:Compression}(a). For our system, $f = 388$, $a = 0.5$ nm, $N_\mathrm{s} = 3410$, and $\nu = 0.53$. The size ratio $\delta = 0.12$, 0.25, 0.35, and 1.36 for linear chains with $M_\mathrm{w} = 150$, 590, 1100, and 15000 kDa, respectively.

According to this theory of how star polymers compress in solutions of linear chains, the size of the PGNPs decreases most drastically in solutions of low $M_\mathrm{w}$ chains (Fig.\ \ref{fig:Compression}(a)). As the free polymer concentration increases, the PGNPs compress. The degree of compression is lower in solutions of higher $M_\mathrm{w}$. These predicted changes in PGNP size are consistent with the conformational changes of the grafted layer elucidated by SANS (Fig. 1 in main text). Because dynamics depend on PGNP size, this compression may be affecting how the PGNP diffusivity depends on free polymer $M_\mathrm{w}$. If the dynamics of PGNPs are solely affected by the PGNP compression, the particles should couple to bulk solution viscosity $\eta$ according to the Stokes-Einstein expression, resulting in
\begin{equation}
\frac{DR}{D_0 R_0} = \frac{\eta_0 }{\eta}
\end{equation}
where $\eta_0$ is solvent viscosity, $R_0$ is the particle size in neat solvent, and $R$ is the compressed size of the PGNP. This ratio, however, fails to collapse the measured PGNP diffusivities, does not agree with the measured bulk rheology, and actually increases the spread in the data (Fig.\ \ref{fig:Compression}(c)). This failure indicates that the PGNP dynamics and predicted compression trend in opposite directions with free polymer $M_\mathrm{w}$. To further exemplify this trend, we calculate the necessary effective size ratio $R_\mathrm{eff}/R_\mathrm{0} = \frac{\eta_0 D_0}{\eta D}$ that would result in the measured diffusivities using the Stokes-Einstein expression and measured solution viscosity. This effective size ratio decreases with increasing polymer concentration but also decreases with increasing free chain $M_\mathrm{w}$, in stark contrast to and opposite from the predicted compression behavior shown in Fig.\ \ref{fig:Compression}. From this analysis, we can conclude that compression of PGNPs cannot explain the observed dynamic behavior. The PGNPs diffuse faster in solutions of high $M_\mathrm{w}$ free polymer \emph{despite} being less compressed.

\begin{figure}[htb!]
\includegraphics[width = 3.25 in ]{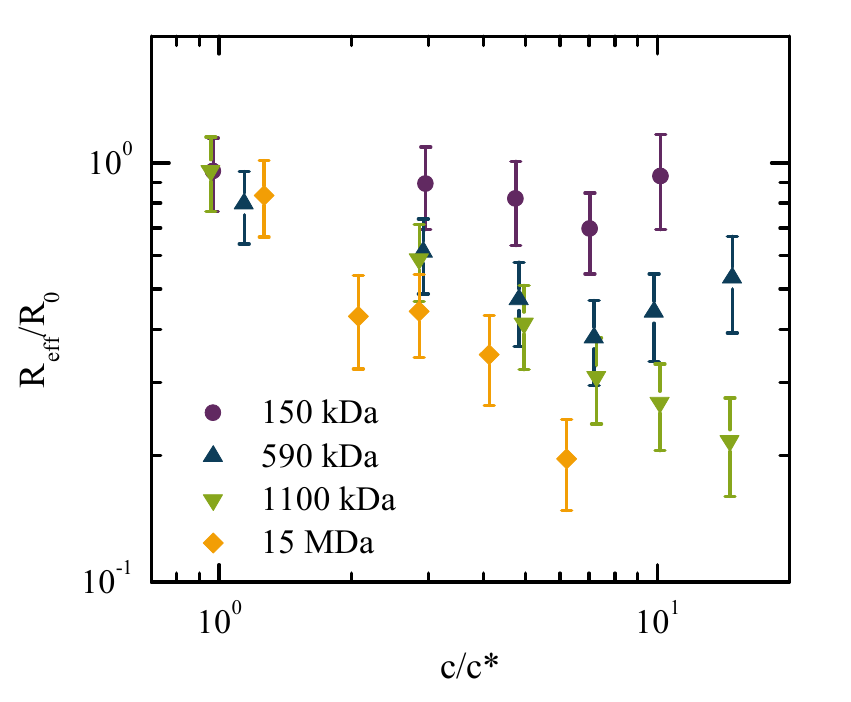}
\caption{\label{fig:Reff} Ratio of effective size of PGNP to the hydrodynamic size in pure solvent $R_\mathrm{eff}/R_\mathrm{0}$ as a function of free polymer concentration for various $M_\mathrm{w}$.}
\end{figure}

\subsection{Dispersion of PGNPs in solutions of linear chains}

\begin{figure}[htb!]
\includegraphics[width = 3.25 in ]{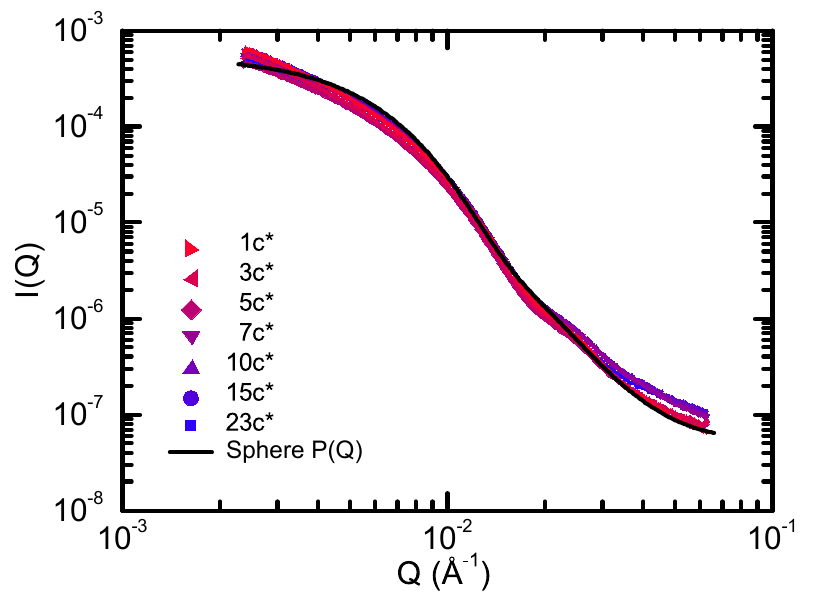}
\caption{\label{fig:SAXS} SAXS intensity $I(Q)$ as a function of wavevector $Q$ for PGNPs dispersed in solution of linear chains with $M_\mathrm{w} = 1100$ kDa at various concentrations.}
\end{figure}

We confirm that the PGNPs remain well dispersed in all solutions using small angle x-ray scattering (SAXS). The scattering from the silica core of the PGNP dominates any scattering from the polymer. Thus, the SAXS intensity $I(Q)$ as a function of wavevector $Q$ is well fit by the hard sphere form factor for a sphere of radius $R = 24$ nm given by 
\begin{equation}
P(Q) = \left(\frac{\sin(QR)-QR\cos(QR)}{(QR)^3}\right)^2
\end{equation}
and a log-normal polydispersity $\sigma = 0.28$ (Fig.\ \ref{fig:SAXS}), in good agreement with our earlier work.\cite{Poling-Skutvik2016, Poling-Skutvik2017} Observing no significant change in $I(Q)$ with increasing polymer concentrations, we conclude that the PGNPs are well-dispersed in solutions of free polymers at all measured concentrations. 

\subsection{Estimates of arm retraction dynamics}

\begin{figure}[htb!]
\includegraphics[width = 3.25 in]{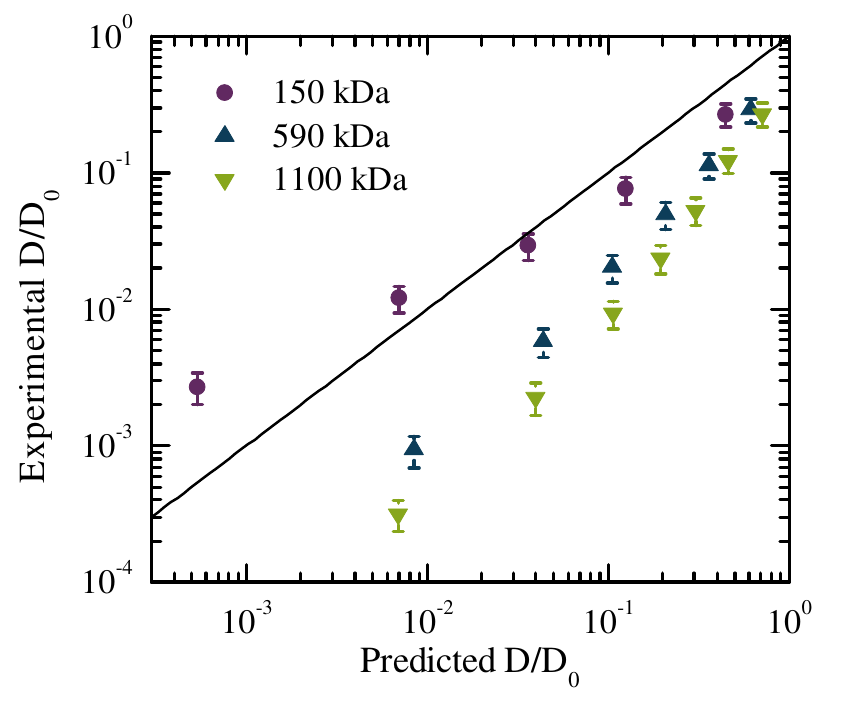}
\caption{\label{fig:ArmRetraction} Experimental diffusivity of PGNPs plotted as a function of predictions based on ref.\ \citenum{McLeish2002} for various free polymer molecular weights. Solid line indicates unity line $y = x$.}
\end{figure}

The dynamics of star polymers in linear solutions are often dependent on the entanglements between arms of the star polymer and surrounding linear chains. These entanglements slow the diffusion of the star core through the linear system. To estimate if similar entanglements affect the movement of the PGNP core, we calculate the number of entanglements per grafted chain using the total dissolved polymer concentration $\phi$ according to\cite{Rubinstein2003}
\begin{equation}
Z = \frac{M_\mathrm{e}}{M_\mathrm{w,g}} = \frac{M_\mathrm{e,0}}{\phi^{4/3} M_\mathrm{w,g}},
\end{equation}
where $M_\mathrm{e,0} = 13000$ Da is the melt entanglement molecular weight of polystyrene.\cite{Lin1987} For all $c/c^*$ and $M_\mathrm{w}$, $Z < 5$. The longest relaxation time of a star polymer is related to $Z$ according to\cite{McLeish2002}
\begin{equation}
\tau_\mathrm{arm} = \frac{\pi^{5/2}}{\sqrt{6}} \tau_\mathrm{e} Z^{3/2} \exp (a Z)
\end{equation}
where $a = 3/2$ is a constant, $\tau_\mathrm{e} = \tau_0 (M_\mathrm{w,g}/M_\mathrm{e})^2 \phi^{-5/3}$ is the relaxation of an entanglement strand, and  $\tau_0 = 6\pi \eta b^3/k_\mathrm{B}T$ is the diffusive relaxation time of a Kuhn segment with $b = 1.54$ \AA.\cite{Rubinstein2003} After $\tau_\mathrm{arm}$, the core of the star should move diffusively with a diffusivity $D_\mathrm{star} = R_\mathrm{H}^2/\tau_\mathrm{arm}$. We plot the measured experimental diffusivities against the predicted normalized values based on this theory derived for star polymers in Fig.\ \ref{fig:ArmRetraction}. Although there is a positive correlation between experiment and the predicted entangled dynamics, there is a significant spread in the data as a function of $M_\mathrm{w}$. Additionally, whereas the diffusivity of PGNPs in low $M_\mathrm{w}$ polymer at high $c/c^*$ are larger than predicted, the diffusivity in high $M_\mathrm{w}$ solutions is much slower than predicted. Further, incorporating tube dilation would lead to faster predicted diffusivities and less agreement between theory and experiments. Thus, entanglements between grafted chains and free linear chains cannot explain the dynamics of PGNPs in semidilute polymer solutions.

\newpage
\bibliography{biblio}